\documentclass[onecolumn]{IEEEtran}
\IEEEoverridecommandlockouts
\usepackage{cite}
\usepackage{amsmath,amssymb,amsfonts,mathcomSTEv4}
\usepackage{graphicx}
\usepackage{textcomp}
\usepackage{xcolor}
\usepackage{cancel}
\usepackage{enumitem}
\usepackage{epstopdf}

\usepackage{graphicx,subfigure,mathcomSTEv4}

\usepackage{tikz}
\usetikzlibrary{patterns,arrows,decorations.pathreplacing}
\usepackage{pgfplots}
\usepackage{circuitikz}

\tikzstyle{joint} = [draw, circle, minimum size=.1em]
\usetikzlibrary{positioning}
\usetikzlibrary{arrows,fit}
\usetikzlibrary{decorations.pathreplacing}
\tikzstyle{int}=[draw, fill=blue!10,thick,minimum height = .75 cm, minimum width=1cm]
\tikzstyle{rbox}=[draw, rounded corners=5 pt,thick ]
\tikzstyle{sum}=[circle, fill=blue!10, draw=black,line width=.5 pt,minimum size = 0.05 cm, thin ]
\tikzstyle{joint} = [draw, circle, minimum size=1em]
\usetikzlibrary{shapes,arrows,backgrounds,positioning}
\usetikzlibrary{positioning,intersections,decorations.pathreplacing}
\tikzstyle{cw}= [fill=blue!10, draw , rounded corners = 1 ex,minimum height = 1cm]
\tikzstyle{joint} = [draw, circle, minimum size=1em]
\tikzstyle{check} = [draw,  minimum size=1em]

\makeatletter

\usepackage[utf8]{inputenc}

\newcommand*{\rom}[1]{\expandafter\@slowromancap\romannumeral #1@}

\makeatother
\def\BibTeX{{\rm B\kern-.05em{\sc i\kern-.025em b}\kern-.08em
   T\kern-.1667em\lower.7ex\hbox{E}\kern-.125emX}}
\begin{document}
\title{On Sum Secure Degrees of Freedom for K-User MISO Broadcast Channel With Alternating  CSIT}

\author{Leyla Sadighi, Sadaf Salehkalaibar, Stefano Rini
	\thanks{Leyla Sadighi and Sadaf  Salehkalaibar are   with the Department of Electrical and Computer Engineering, College of Engineering, University of Tehran, Tehran, Iran (e-mails:  {\{leyla.sadighi,s.saleh\}@ut.ac.ir})}
	\thanks{Stefano Rini is with the Department of Electrical and Computer Engineering, National Chiao Tung University, Taiwan  (e-mail: stefano@nctu.edu.tw).}
}

\maketitle

\begin{abstract}
In this paper, the sum secure degrees of freedom (SDoF) of the  $K$-user Multiple Input/Single Output (MISO)  Broadcast Channel with Confidential Messages (BCCM) and alternating Channel State Information at the Transmitter (CSIT) is investigated.

In the MISO BCCM, a $K$-antenna transmitter (TX) communicates toward $K$ single-antenna receivers (RXs), so that message for RX $k$ is kept secret from  RX $j$ with $j<k$.  
For this model, we consider the scenario in which the CSI of the RXs from $2$ to $K$ is instantaneously known at the transmitter while CSI of RX $1$ is known at the transmitter (i) instantaneously for half of the time and (ii) with a unit delay for the remainder of the time. 
We refer to this CSIT availability as \emph{alternating} CSIT.
Alternating CIST has been shown to provide synergistic gains in terms of SDoF and is thus of a viable strategy to ensure secure communication by simply relying on the CSI feedback strategy.
Our main contribution is the characterization of sum SDoF for this model as $SDoF_{\rm sum}= (2K-1)/2$. 
Interestingly, this $SDoF_{\rm sum}$ is attained by a rather simple achievability in which the TX uses artificial noise to prevent the decoding of the message of the unintended receivers at RX $1$.
For simplicity first, the proof for the case $K=3$ is discussed in detail and after that, we have presented the results for any number of RXs. 
\end{abstract}

\begin{IEEEkeywords}
Broadcast channel with confidential messages; Multiple input/single output channel;  Secrecy capacity; Degrees of freedom; Alternating channel state information.
\end{IEEEkeywords}	

\section{INTRODUCTION}

The Broadcast Channel with Confidential Messages (BCCM) is the multi-terminal channel in which one transmitter (TX) communicates toward a set of receivers (RX) so that the message of one user remains secret from a given set of receivers.  
In this paper, we study the  Multiple Input/Single Output (MISO) version of this channel: a transmitter with $K$ antennas communication toward $K$ receivers with a single antenna over an Additive White  Gaussian Noise (AWGN) channel. 
This channel model is particularly relevant in modern wireless communication scenarios in which illegal eavesdropping of down-link communications is easily accomplished.
For this scenario,  we leverage channel state information at the transmitter (CSIT) to guarantee private and secure communication, that is: since the channel realization between the TX and each of the RX is unknown at the other RXs, this source of randomness can be used to achieve secure communication. 
As having the RXs feedback the CSI to the TX is expensive in terms of energy and computational complexity, one would want to minimize the CSIT availability so as to satisfy the security demands of each of the RX. 
For this reason, 
we investigate the high-SNR asymptotic of the secure communication performance attainable through alternating CSIT in the form of the Sum Secure Degrees of Freedom ($SDoF_{\rm sum}$). 
Our results, although theoretical in nature, validate the effectiveness of a particularly simple strategy to attain secrecy: transmitting artificial noise toward the non-intended receiver to obfuscate the messages for other intended receivers. 
In the following, we indicate the CSIT availability as a vector with entries $P$ or $D$ to indicate whether the CSIT is available perfectly, delayed respectively. 

\subsubsection*{Literature review}
Let us briefly review the literature on the SDoF of multi-terminal channels, such as the Broadcast Channel with Confidential Messages (BCCM),  also relying on alternating CSIT.
In \cite{awan2016sdof}, the authors considered the problem of secure transmission over a $2$-user MISO broadcast channel with an external eavesdropper. 
First, they characterized the SDoF region of fixed CSIT states PPD, PDP, and DDP for the first RX, second RX, and eavesdropper respectively. 
Next, the authors established bounds on the SDoF region on the symmetric case in which the transmitter is allowed to alternate between PDD and DPD states with equal fractions of time.
When considering more than two receivers,  most literature has focused on the MISO case in which the number of transmit antennas equals the number of RXs.  
%
%
%
The SDoF of a $3$-user MISO BCCM when the channel state alternates between the PPP and DPP states at the RXs with an equal fraction of time is investigated in \cite{alipour2017secure}.

For the BC with a secrecy constraint, which is the channel in which the message for one RX has to be kept secret from all other RXs,
%
%
  partial (Perfect CSIT for some users and Delayed for the others) results are presented for the multi-user MISO BC with $M$ transmit antennas and $K$ single-antenna users in \cite{mukherjee2015secrecy}. For this problem, it is characterized that the minimum amount of perfect CSIT required per user to achieve the maximum DoFs of $ \min(M,K) $  is $ \min(M,K)/K $.
%
%
The DoF for the $K$-user MISO BC with alternating CSIT is analyzed in \cite{seif2016achievable}, and total achievable DoF is given by $\frac{K^2}{2K-1}$. 
%

\subsubsection*{Contributions}
In this work, we determine the $SDoF_{\rm sum}$ for the MISO BCCM with $K$ receivers and specific secrecy constraints where the TX with $K$ antennas transmit toward $K$ RXs each with one antenna, in such a way that the message for RX $k$ is kept secret from RX $j$ for all $j<k$, as shown in Fig.~\ref{fig1}. 
 As such, the present work is an effort to define different levels of confidentiality in the model of BC at the high-SNR regime.
Indeed, when considering a practical communication scenario, messages have different significance that correspond to different confidentiality levels. 
For this reason, we define the secrecy constraints in a way such that messages with higher importance have higher levels of confidentiality. 
%
For example, for the case $K=3$, RX $1$ is the one who is not security conscious thus only feedbacks half of its CSI to the TX, RX $3$ is the most security-conscious, so it provides full CSI to the TX. RX $2$ is partially security conscious 
but still feedbacks all its CSI.
%
The strength of our work compared to previous works is in extending the analysis to the model with any number of users, which, in terms,  also has stronger confidentiality constraints. 
Both the achievability and the converse proofs rely on the synergic benefits of alternating CSIT to achieve optimal $SDoF_{\rm sum}$.

\subsubsection*{Paper Organization}
The remainder of the paper is organized as follows. In section \rom{2}, we will first present our system model and a mathematical framework. The relevant results for $2$-user, $3$-user, and $K$-user BC are mentioned in section \rom{3}. Our main result is presented in Section \rom{4}, and the proofs of the main result are provided in \rom{5}. We finally conclude the paper in Section \rom{6}.
\subsubsection*{Notation}
With the notation $[n:m]$ indicates the set $\{n,n+1, \ldots, m-1,m\}$. We also adopt the shorthand $[n] \triangleq [1:n]$. 
The  variable of the $i^{\rm th}$ receiver is indicated with the subscript $i$, i.e. $X_i$. 
The time-dependency is indicated in brackets, i.e. $X(t)$. 
We also adopt the short hand notation $\{ x(t)\}_{t \in [n]}$  as $x^n$.
Vectors are indicated using bold lower-case letters, i.e., $\vv$, all vectors are taken to be column vectors.
Matrices are indicated using bold upper-case letters, i.e., $\Mv$.
Random variables/vectors (RVs) are indicated with upper case letters, i.e., $X$.
With $\Xcal$, we indicate the support of the RV $X$.
The notation $\Ccal \Ncal( \muv , \Sigma )$ indicates the circularly symmetric Gaussian distribution with mean $\muv$ and covariance matrix $\Sigma$.

\section{SYSTEM MODEL AND DEFINITIONS}
\label{sec:SYSTEM MODEL AND DEFINITIONS}

\begin{figure}
	\centering
	\hspace{-0.5 cm}
	\begin{tikzpicture}[thick,scale=1.3, every node/.style={scale=0.9}]
		%
		\node at (1,3) (source) {} ;
		
		\node [int, right of = source, node distance = 3.5 cm](enc) {Enc.};

		\node at (8,6) [int, node distance = 1.75 cm](dec1){\large Dec. 1};
		\node  at (8,4) [int, node distance = 1.75 cm](dec2){Dec. 2};
		\node   at (8,2)[ node distance = 1.75 cm](decDots){$\vdots$};
		\node   at (8,0)[int, node distance = 1.75 cm](decK){Dec. K};
		
		\node [right of = dec1, node distance = 3.5 cm](end1){};
		\node [right of = dec2, node distance = 3.5 cm](end2){};
		\node [right of = decK, node distance = 3.5 cm](endK){};
		
		
		\draw[->,line width=1pt] (source) -- node [above] {  $W_1, W_2 \ldots W_K$} (enc);
		\draw[->,line width=1pt] (dec1) -- node [above] {  $W_1, \cancel{W_2} \ldots \cancel{W_K}$} (end1);
		\draw[->,line width=1pt] (dec2) -- node [above] {  $W_2, \cancel{W_3} \ldots \cancel{W_K}$} (end2);
		\draw[->,line width=1pt] (decK) -- node [above] {  $W_K$} (endK);

		\node [draw,  rectangle, left of = dec1, node distance = 2.5 cm](h1){$\hv_1(t)$};
		\node [draw,  rectangle, left of = dec2, node distance = 2.5 cm](h2){$\hv_2(t)$};
		\node [ left of = decDots, node distance = 2.5 cm](h3){$\vdots$};
		\node [draw,  rectangle, left of = decK, node distance = 2.5 cm](hK){$\hv_K(t)$};
		
		\node [joint, right of = h1, node distance = 1.25 cm](Z1){};
		\node [joint, right of = h2, node distance = 1.25 cm](Z2){};
		\node [joint, right of = hK, node distance = 1.25 cm](ZK){};
		
		\node [above of = Z1, node distance = 1 cm](Z11){$N_1(t)$};
		\node [above of = Z2, node distance = 1 cm](Z12){$N_2(t)$};
		\node [above of = ZK, node distance = 1 cm](Z1K){$N_K(t)$};
		
		\node [right of = Z1, node distance = 0 cm]{$+$};
		\node [right of = Z2, node distance = 0 cm]{$+$};
		\node [right of = ZK, node distance = 0 cm]{$+$};    

		\draw[<-,line width=1 pt] (Z1) -- (Z11);
		\draw[<-,line width=1 pt] (Z2) -- (Z12);
		\draw[<-,line width=1 pt] (ZK) -- (Z1K);

		\draw[->,line width=1 pt] (enc)|-(h1);
		\draw[->,line width=1 pt] (enc)|-(h2);
		\draw[->,line width=1 pt] (enc)|-(hK);

		\draw[<-,line width=1 pt] (Z1)--(h1);
		\draw[<-,line width=1 pt] (Z2)--(h2);
		\draw[<-,line width=1 pt] (ZK)--(hK);

		\draw[->,line width=1 pt] (Z1)--(dec1);
		\draw[->,line width=1 pt] (Z2)--(dec2);
		\draw[->,line width=1 pt] (ZK)--(decK);
		
		\node at (3.7,8) [](switch1){CSI $\hv_1(t)$};
		\node at (4.35,7) [spdt,rotate=180]{};
		
		\node at (3.5,7.25) [draw] (delay){D};
		
		\draw[-,dotted,line width=1 pt] (h1.north west)|-(4.5,7);
		\draw[-,dotted,line width=1 pt] (h2.south west) --(hK.north west);
		\draw[-,dotted,line width=1 pt] (enc.north west)|-(delay);
		\draw[<-,dotted,line width=1 pt] (enc.north west)|-(4,6.675);
		
		\node at (4,-1) (low_bend){};
		\draw[-,dotted,line width=1 pt] (hK.south west) |-(low_bend.west);
		\draw[<-,dotted,line width=1 pt] (enc.south west) |-(low_bend.west);
		
		\node at (4.75,-1.5) [] {CSI $\{\hv_2(t), \ldots, \hv_K(t)\}$};
		
	\end{tikzpicture}
	
	\caption{$K$-user Multiple Input/Single Output (MISO)  Broadcast Channel with Confidential Messages (BCCM) and alternating Channel State Information at Transmitter (CSIT) which alternates from $P^K$ channel state to $DP^{K-1}$ }
	\label{fig1}
\end{figure}
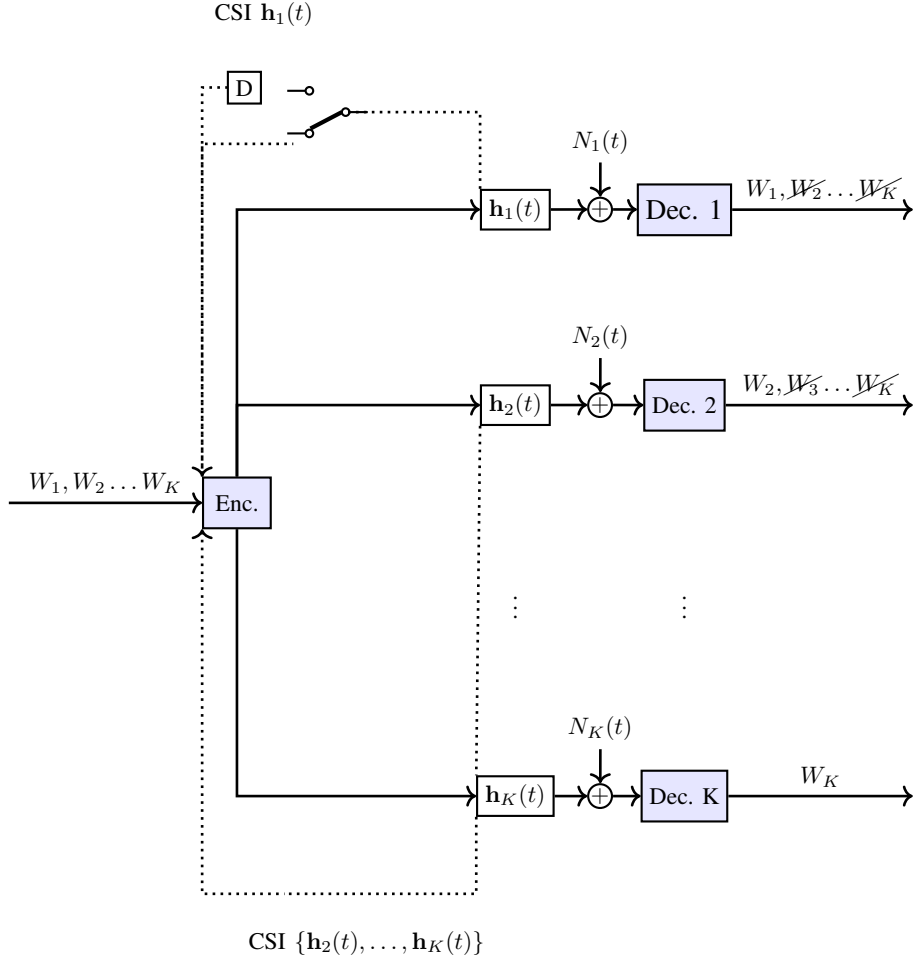

A $K$-user  Broadcast Channel with Confidential Messages (BCCM), consists of a $K$-user broadcast channel (BC) in which some messages are shared between users while the others should be concealed from unintended receivers based on secrecy conditions.
The $K$-user BCCM and with alternating CSIT is a variation of the BCCM in which the CSIT is provided in an alternating fashion.
In the following, we consider the case in which (i) half of the time, the CSI of the first receiver is known perfectly at the transmitter, while (ii) the other half the time this CSI is known with a delay of one time unit.
The CSI of the other receivers is always perfectly known at the transmitter. 

More specifically, we consider the multiple input/single output (MISO) BCCM in which the transmitter (TX) is equipped with $K$ antennas, while each of the $K$ receivers (RX) is equipped with one antenna, as depicted in Fig.~\ref{fig1}. 
The transmitter communicates to the receivers over $T$ channel uses. 
 The input/output relationship between the transmitter and  the {$k^{\rm th}$} receiver at time instant $t \in [T]$ is obtained as
\begin{equation}\label{I/O}
Y_{k}(t) = \hv_{k}^{\mathsf{H}}(t) X(t)+N_{k}(t),  \quad \forall \ k\, \in [K], 
\end{equation}
where $ X(t) \in \Cbb^{K} $ is the channel input (column) vector, $Y_{k}(t) \in \Cbb$ is the channel output, $\hv_{k}(t) \in \Cbb^{K}$, is the channel state vector and $N_k(t) \in \Cbb$, $N_k(t) \sim \Ccal\Ncal(0,1)$ is the AWGN.
Each entry in the channel state vector is obtained as i.i.d. drawn from the continuous distribution of $P_H$. 
Additionally, the channel input is subject to the second moment constraint 
\ea{
	\sum_{t \in [T]}\Ebb \lsb \left\| X(t)\right\|_2^2 \rsb \leq TP,
	\label{eq:power constraint}
}
where $\| \cdot \|_2$ is the $L_2$ norm and $P \in \Rbb^+$. 
When vectorizing the channel output over the user index $k \in [K]$, we obtain the more compact expression 
\ea{
	Y(t)=\Hv(t) X(t) + N(t),
	\label{eq:vectorize}
}
where we have used the vectorization 
\ea{
	X(t)=[X_1(t), \ldots, X_K(t)],
	\label{eq:vectorize}
}  for ${X(t)}, \ Y(t)$ and $N(t)$ while 
\ea{
	\Hv(t)= \lsb \hv_1(t), \hv_2(t), \ldots, \hv_k(t)  \rsb^{\mathsf{H}}.
}
In the following, we use the notation
\ea{
	\Scal^t=\lcb \Hv(i) \rcb_{i \in [t]},
}
to compactly indicate the CSI of all users up to time $t$.
The channel state information (CSI) is assumed to be made available at the transmitter in the following fashion: 
\begin{enumerate}[label=(\roman*)]
	\item {\bf Perfect CSIT for RX $[2:K]$: }    
	the CSI $\{\hv_2(t), \ldots, \hv_K\}$ is available to the transmitter at time $t$,
	\item {\bf Perfect CSIT for RX $1$ for half of the channel uses:}  if $t$, i.e. $t \in [1, T/2 ]$, the CSI $ \hv_1(t)$ is instantaneously available, at the transmitter,
	\item {\bf Delayed CSIT for RX $1$ for half of the channel uses:}  if $t$, i.e. $t \in [T/2, T]$, the CSI $ \hv_1(t)$ is made available at the transmitter with a unit time delay. 
\end{enumerate}
%
We refer to this CSIT availability  above as \emph{alternating} CSIT \cite{tandon2013synergistic} and we  indicate in Fig. \ref{fig1} as a switch. 
%
%
Finally, we assume that the CSI availability that is whether the CSI is perfect or delayed, is known at all users instantaneously.

Next, we introduce some standard definitions of code, achievable region, and SDoF.
In the MISO BCCM,
the TX wishes to communicate the message  $\Wcal_k=\lsb 2^{\lfloor n R_k \rfloor}\rsb$ to user $k$.
A code for the $K$-user MISO BCCM with alternating CSIT consists of the two encoding mappings.
\eas{
f_{{\rm enc}, P}(t, \cdot ) & :\prod_{k \in [K]}  \Wcal_k \times  \Scal^{t} \goes \Xcal \\
f_{{\rm enc}, D}(t, \cdot ) & :\prod_{k \in [K]}  \Wcal_k \times  \Scal^{t-1} \prod_{i \in [2:K]} \Hcal_i \goes \Xcal,
}
($P$ for perfect CSIT from RX $1$, $D$ for delayed CSIT from RX $1$) and  $K$ decoding mappings
\ea{
f_{ {\rm dec}, k}(t, \cdot) :  \Xcal^t \times \Scal^{t-1} \times \Hcal_k(t)  \goes  \Wcal_k.
}
For such a code, the probability of error at time $T$, $P_e(T)$, defined as
\ea{
P_e(T)=\max \ \Pr \lsb f_{ {\rm dec}, k}(T, Y_k^T, \Scal^{T-1},\hv_k(T))  \neq W_k \rsb,
\label{eq: max Pe}
}
where $Y_k^T$ is obtained by applying the encoding functions  
\ean{
& \lcb f_{{\rm enc}, P}\lb t, [W_1, \ldots, W_k], \Scal^t \rb \rcb_{t \in [T/2]}, \  \\
& \lcb f_{{\rm enc}, D}\lb t, [W_1, \ldots, W_k], \Scal^{t-1},[\hv_2, \ldots, \hv_K]\rb \rcb_{t \in [T/2:T]},
}
to produce the sequence of inputs $\{X(t)\}_{t \in [T]}$\footnote{Note that the maximum in \eqref{eq: max Pe} is over all messages and all users. Also note that the dependency of the functions from the power constraint in \eqref{eq:power constraint} is not indicated explicitly. }.c

A rate tuple $[R_1, R_2, \ldots,R_K]$ is said to be \emph{securely achievable} if there exists a sequence of codes such that the probability of error vanishes at $T \goes \infty$
while also 
\ea{
	I \lb W_k;\left\lbrace Y_j^T \right\rbrace_{0\leq j \leq k-1} |\Scal^T \rb \leq T \epsilon , \qquad     \forall \ k\in [K],
	\label{confidentiality}
}
In the literature, the condition in \eqref{confidentiality} is referred to as \emph{perfect secrecy}.

The \emph{secure capacity region}  is the convex hull of all the securely achievable rates. 
The \emph{sum secure capacity}, denoted as $\Ccal_{\rm sum}(P)$ of the BCCM with alternating CSIT is the achievable rate tuple attaining {the supremum} of the sum of the rates $\{R_k\}_{k \in[K]}$, i.e. $R_{\rm sum}=\sum_{k \in [K]} R_k$.
Finally, the SDoF is obtained as 
\ea{
	SDoF_{\rm sum}= \lim_{P \goes \infty} \f{\Ccal_{\rm sum}(P)}{ \log (P)}.
}
\section{Relevant Results}
Let us briefly review some results on the SDoF of various BCCM channels as available in the literature.
\subsection{$3$-user MISO BCCM}
In \cite{mohanty2013dof}, for a $3$-user MISO BC, when the transmitter has perfect CSIT of the channel to one receiver and delayed CSIT of channels to the other two receivers, two new communication schemes are proposed that can achieve a DoF tuple $ (1,\frac{1}{3},\frac{1}{3}) $ with a $DoF_{\rm sum}$ of $\frac{5}{3}$. 
The SDoF is, instead, studied in \cite{awan2016sdof}.
\begin{thm}{\bf \cite{awan2016sdof} } \label{r1}
	For the $3$-user MISO BCCM, the SDoF with either perfect or delayed CSIT is obtained as
	\ean{
		& SDoF_{\rm sum}^{PPD}=2, \quad 		SDoF_{\rm sum}^{PDP}= \f 3 2, \\
		& SDoF_{\rm sum}^{DDP}=\f 4 3, \quad 	SDoF_{\rm sum}^{PDD}=1 .
	}
\end{thm}
Additionally, in \cite{awan2016sdof},  the authors establish bounds on the SDoF region on the symmetric case in which the transmitter is allowed to alternate between PDD and DPD states in equal fractions of time. 
In \cite{buzuverov2019dof}, the optimal DoF for the 2-antenna $3$-user MISO BC with alternating CSIT where the permitted CSIT states are PPP, PPD, PDP, PDD, and DDD is characterized.
\subsection{$K$-user MISO BCCM}
In \cite{tandon2013synergistic},  partial results are presented for the multi user MISO BC with $M$ transmit antennas and $K$ single-antenna users. For this problem, it is characterized that the minimum amount of perfect CSIT required per user to achieve the maximum DoFs of $\min(M,K) $  is $\min(M,K)/K $.
The work in \cite{seif2016achievable} studied the DoF of the $K$-user MISO BC through utilizing the synergistic benefits of alternating CSIT. The authors consider perfect, delayed, and unknown CSIT for each user among different time slots. They calculated the distribution for a fraction of time at each state as
\begin{equation*}
\lambda_{P}=\frac{(K-1)^2}{2K^{2}-K}, \quad \lambda_{D}=\frac{K-1}{2K-1}, \quad \lambda_{N}=\frac{1}{K},
\end{equation*}
and showed that the achievable DoF for the proposed network is given by $DoF_{\rm sum}=\frac{K^2}{2K-1}$.
\section{MAIN RESULT}

The main result of the paper is shown by the next theorem. 
\begin{thm}{\bf $K$-user SDoF.}
	\label{th:main theorem 1}
	For the $K$-user MISO BCCM with alternating CSIT in Fig \ref{fig1}, we have:
	\ea{
		SDoF_{\rm sum}=\frac{2K-1}{2}.
	}

\end{thm}
Because of the simplicity of understanding, we will first present the case of $K=3$ users in detail in Sec. \ref{Sec. 3-User $PPP/DPP$ BCCM Channel Case} and then we will move on the case of any $K$ is Sec. \ref{Sec. K-User $D1/P1$  BCCM Channel Case}.
\subsection{3-User $PPP/DPP$ BCCM Channel}
\label{Sec. 3-User $PPP/DPP$ BCCM Channel Case} 
Let us next present in some detail the proof of the following corollary of Th. \ref{th:main theorem 1}. 

\begin{cor}{\bf $3$-user SDoF.}
	\label{cor:main th}
	For the $3$-user MISO BCCM with alternating CSIT  in Sec. \ref{sec:SYSTEM MODEL AND DEFINITIONS}, we have 
	\ea{
		SDoF_{\rm sum}=\frac{5}{2}.
		\label{eq:main th}
	}
\end{cor}
\begin{IEEEproof}
	The proof is divided into achievability and converse parts. 
	Conceptually, the proof will show that in the $PPP$ state, it is possible to attain $SDoF_{\rm sum}=3$, since the transmitter can use orthogonal transmissions toward each receiver.
	In the state $DPP$, instead, only $SDoF_{\rm sum}=2$ can be achieved, since orthogonal transmissions {cannot be used to hide the messages of RX $2$ and RX $3$ from RX $1$.}
	In this latter case,  artificial noise is transmitted toward RX $1$, so to hide these messages of RX $2$ and RX $3$.
	This artificial noise is orthogonal to the signal space at RX $2$ and RX $3$,  to allow secure transmissions toward these two receivers. 
	Note that the specific order in which the states $PPP$ and $DPP$ occur does not impact the SDoF since the state is known at all users.
	Also note that, for the case of three receivers, the perfect secrecy conditions in \eqref{confidentiality} are obtained as
	
\ea{
	\label{confidentiality 3 users 1,2} 
	I \lb W_2; Y_1^T  |\Scal^T \rb  \leq T \epsilon \, ,
	I \lb W_3; Y_1^T, Y_2^T  |\Scal^T \rb  \leq T \epsilon ,
}	
	\smallskip
	\noindent
	$\bullet$ \emph{Achievability:}
	As we are concerned with the high SNR asymptotics, we will disregard the effect of the additive noise.
	The achievability proof is as follows: without loss of generality, assume that the channel input at time $t=t_{\rm P}\in [1 , \lfloor T/2 \rfloor]$ is in the state $PPP$ and in state $DPP$ at time $t=t_{\rm D} \in [ \lfloor T/2 \rfloor,T ]$.
	For each pair  of times $[t_D , t_P]$, we send one symbol, $a_{1}$ to RX $1$ and two symbols, i.e., $ \left(a_2,b_2 \right) $ and $ \left(a_3,b_3 \right) $, to  RX $2$  and RX $3$. This is accomplished as follows.
	Let $X(t_{\rm P})$ be obtained as 
	\begin{align}
	X(t_{\rm P})= a_{1}\vv_1(t_{\rm P})+a_2\vv_2(t_{\rm P})+a_3\vv_3(t_{\rm P}),
	\end{align}
	%
where $\vv_k(t_{\rm P})$, $k \in [3]$ are vectors satisfying the following linear independence conditions
	\ea{ 
		\left\langle \hv_2(t_{\rm P}), \vv_1(t_{\rm P})\right\rangle  & =\left\langle {\hv}_3 (t_{\rm P}) ,\vv_1(t_{\rm P})\right\rangle =0  \nonumber  \\ 
		\left\langle \hv_1 (t_{\rm P}) ,\vv_2(t_{\rm P})\right\rangle  & =\left\langle {\hv}_3(t_{\rm P}), \vv_2(t_{\rm P})\right\rangle =0 \nonumber  \\
		\left\langle \hv_1 (t_{\rm P}),\vv_3(t_{\rm P})\right\rangle  & =\left\langle \hv_2 (t_{\rm P}), \vv_3(t_{\rm P})\right\rangle =0,
	}
	
so that the symbol $a_k$ can be securely decoded at RX $k$.
	At the time instant $t=t_{\rm D}$, the encoder can no longer securely communication to RX $1$ using interference nulling: in this case we let the channel input be
	\ea{
		X(t_{\rm D})= U \vv_1(t_{\rm D})+b_2\vv_2(t_{\rm D})+b_3 \vv_3(t_{\rm D}),
		\label{eq:X(t D)}
	}
	where, this time we satisfy the orthogonality conditions 
	\ea{
		\left\langle \hv_2(t_{\rm D}) ,\vv_1(t_{\rm D})\right\rangle =\left\langle \hv_3(t_{\rm D}),\vv_1(t_{\rm D})\right\rangle =0   \nonumber \\ 
		\left\langle \hv_2(t_{\rm D}), \vv_3(t_{\rm D})\right\rangle =\left\langle \hv_3(t_{\rm D}),\vv_2(t_{\rm D})\right\rangle =0,
	}
	while $U \sim \Ccal \Ncal(0,P)$  is an artificial noise aimed at concealing the private messages $b_2$ and $b_3$ from RX $1$.
	With the choice of channel input in \eqref{eq:X(t D)}, we obtain the channel outputs 
	\ea{
		Y_1(t_{\rm D})& = U \left\langle \hv_1(t_{\rm D}), \vv_1(t_{\rm D})\right\rangle + b_2 \left\langle \hv_1(t_{\rm D}) ,\vv_2(t_{\rm D})\right\rangle  \nonumber \\
		& \quad \quad +  b_3 \left\langle \hv_1(t_{\rm D}) ,\vv_3(t_{\rm D})\right\rangle    \triangleq L(b_2,b_3,U) \nonumber  \\
		Y_2(t_{\rm D}	)& =b_2\left\langle \hv_2(t_{\rm D}),\vv_2(t_{\rm D})\right\rangle  \nonumber  \\
		Y_3(t_{\rm D})& =b_3\left\langle \hv_3(t_{\rm D}),\vv_3(t_{\rm D})\right\rangle .
	}
	Since $5$ messages where transmitted in $2$ time instants, we have that $SDoF_{\rm sum}\geq \dfrac{5}{2}$.
	To show that the SDoF of $\frac{5}{2}$ is indeed achievable, it is necessary to verify the constraints in \eqref{confidentiality 3 users 1,2}. 
	Let us begin the secrecy constraint for the message for RX $2$ in the signal received at RX $1$:
	\ean{
		I(W_2; Y_1^T|\Scal^T)&=\f T 2 I(a_2,b_2; a_1 \left\langle \hv_1(t_{\rm P}),\vv_1\right\rangle ,L(b_2,b_3,U)|\Scal^T)\\
		& \leq  T \Ocal(\log P),
	}
	where we have used the fact that the message symbols are independent and the fact that, due to the high-power artificial noise $U$, it is impossible to reliably obtain the linear combination of $a_2$ and $b_2$ from $Y_1$.
	Next, let us consider the secrecy of the message for RX $3$ in the signal received at RX $1$ and RX $2$:
	\ean{
		& I(W_3; Y_1^T,Y_2^T|\Scal^T) = \\
		& \f T 2 I(a_3,b_3; a_1 \left\langle \hv_1(t_{\rm P}),\vv_1\right\rangle ,a_2 \left\langle \hv_2(t_{\rm P}),\vv_2\right\rangle , \\ 
		& \quad \quad b_2 \left\langle \hv_2(t_{\rm D}),\vv_2(t_{\rm D})\right\rangle ,L(b_2,b_3,U)|\Scal^T)\\
		& \quad \leq  T \Ocal (\log P),
	}
	where, again, we have used the independence of the message symbols and the effect of the artificial noise. \\
	\smallskip
	\noindent
	$\bullet$ \emph{ Converse:}
	For compactness of notation, let us indicate $\{Y_i(t)\}_{t \in [1 , \lfloor T/2 \rfloor]}$ as $Y_{iD}^M$ and  $\{Y_i(t)\}_{t \in [ \lfloor T/2 \rfloor,T ]}$ as $ Y_{iP}^M$ for $M=\lfloor T/2 \rfloor$.
The converse hinges on the lemma \ref{lem_sep1}. The proof of this lemma is in Appendix \ref{app:lem_sep2} for any arbitrary number of $K$  users. In this part, we use the corollary of this lemma for three users as follow: 
\begin{cor}{\bf $3$-user SEP.}
	\label{eq:doubling MI}
	In the $3$-user MISO BCCM channel with alternating CSIT we have
	\begin{align}\label{conv4}
		&2H({Y}_{1}^{T}| Y_{2}^{T} , W_{1},\Sbb^T) \geq 
		H({Y}_{1}^{T},Y_{3D}^M|Y_{2}^{T} , W_{1},\Sbb^T).
	\end{align}
\end{cor}
Next, we  write
\eas{
	\label{eq:0}
	& T(R_1 + R_2 +R_3 ) \\
	& \quad \quad =  H( W_{1}|\Sbb^T)+H( W_{2}|\Sbb^T)\nonumber\\
	& \quad \quad  \quad+ H( W_3|W_{1},\Sbb^T) \nonumber \\
	&\quad \quad  \leq I(W_1;Y_{1}^{T}|\Sbb^T)+  
		I(W_2;Y_{2}^{T}|\Sbb^T)\nonumber\\
	&\quad \quad \quad  + I(W_3;Y_3^{T}|W_1,\Sbb^T)+3T\epsilon  
	\label{eq:2} \\
	&\quad \quad  \leq I(W_1;Y_1^T|\Sbb^T)+I(W_2;Y_2^T|Y_1^T,\Sbb^T)\nonumber\\
	& \quad \quad  \quad +	I(W_3;Y_3^T|Y_1^T,Y_2^T,W_1,\Sbb^T)+6T\epsilon, 
	\label{eq:3}
}
where \eqref{eq:2} follows by Fano’s inequality, and  \eqref{eq:3} follows from the following inequality
\ea{
	& I(W_2;Y_{2}^{T}|\Sbb^T) \nonumber \\
	& \quad \leq I(W_2;Y_{1}^T,Y_{2}^T|\Sbb^T) \nonumber \\
	& \quad =I(W_2;Y_{1}^T|\Sbb^T)\nonumber+I(W_2;Y_2^T|Y_{1}^T,\Sbb^T)+\ep T \\
	& \quad \leq  I(W_2;Y_2^T|Y_{1}^T,\Sbb^T) +T\epsilon\label{eq0},
} 
where, in \eqref{eq0}, we have used the secrecy condition in \eqref{confidentiality 3 users 1,2}.
In \eqref{eq:3}, we also use the following bound
\eas{
	&  I(W_3;Y_3^{T}|W_1,\Sbb^T)  \nonumber  \\
	& \quad \quad \leq I(W_3;Y_{1}^T,Y_{2}^T,Y_3^{T}|W_1,\Sbb^T)\\
	& \quad \quad \leq I(W_3;W_1,Y_{1}^T,Y_{2}^T,Y_3^{T}|\Sbb^T)
	\label{eq:a1}\\
	& \quad \quad =I(W_3;Y_{1}^T,Y_{2}^T,|\Sbb^T)\nonumber\\
	&\quad \quad \quad +I(W_3;W_1|Y_{1}^T,Y_{2}^T,\Sbb^T) \nonumber\\
	& \quad \quad \quad \quad +I(W_3;Y_3^T|Y_{1}^T,Y_{2}^T,W_1,\Sbb^T)\\
	&\quad \quad \leq I(W_3;Y_3^T|Y_{1}^T,Y_{2}^T,W_1,\Sbb^T)\nonumber\\
	& \quad \quad \quad +H(W_1|Y_{1}^T,Y_{2}^T,\Sbb^T)\nonumber \\
	&\quad \quad \quad -H(W_1|Y_{1}^T,Y_{2}^T,W_3,\Sbb^T)+T\epsilon\label{eq:a2}\\
	& \quad \quad \leq I(W_3;Y_3^T|Y_{1}^T,Y_{2}^T,W_1,\Sbb^T)\nonumber\\
	& \quad \quad \quad + 	H(W_1|Y_{2}^T,\Sbb^T)+T\epsilon\\
	& \quad \quad \leq I(W_3;Y_3^T|Y_{1}^T,Y_{2}^T,W_1,\Sbb^T) + 2T\epsilon\label{eq:a3},
} 
where \eqref{eq:a1} is due to the independency of messages, \eqref{eq:a2} follows from the secrecy conditions in \eqref{confidentiality 3 users 1,2}	, and \eqref{eq:a3} from Fano’s inequality.
An alternative bound on the sum rate is obtained as 
\eas{
	& T(R_1 + R_2 +R_3 ) \\
	& \quad  \quad  \leq H(Y_1^T|\Sbb^T)-H(Y_1^T|W_1,\Sbb^T) 	\nonumber\\
	&\quad  \quad  \quad  +H(Y_2^T|Y_1^T,\Sbb^T)-H(Y_2^T|W_2,Y_1^T,\Sbb^T)\nonumber\\
	&\quad  \quad  \quad   +H(Y_3^T|Y_1^T,Y_2^T,W_1,\Sbb^T)\nonumber\\
	& \quad  \quad  \quad  -H(Y_3^T|Y_1^T,Y_2^T,W_1,W_3,\Sbb^T)+6T\epsilon \\
	& \quad  \quad  = H(Y_1^T|\Sbb^T)-H(Y_1^T|W_1,\Sbb^T)\nonumber\\
	& \quad  \quad  \quad  	+H(Y_2^T|Y_1^T,\Sbb^T)-H(Y_2^T|W_2,Y_1^T,\Sbb^T)\nonumber\\
	& \quad  \quad  \quad  +H(Y_{3D}^M,Y_{3P}^M|Y_1^T,Y_2^T,W_1,\Sbb^T)\nonumber\\
	& \quad  \quad \quad    -H(Y_3^T|Y_1^T,Y_2^T,W_1,W_3,\Sbb^T)+6T\epsilon \\
	& \quad  \quad   \leq H(Y_1^T|\Sbb^T)-H(Y_1^T|W_1,\Sbb^T)\nonumber\\
	& \quad  \quad  \quad  	+H(Y_2^T|Y_1^T,\Sbb^T)-H(Y_2^T|W_2,Y_1^T,\Sbb^T)\nonumber\\
	& \quad  \quad  \quad  +H(Y_{3D}^M|Y_1^T,Y_2^T,W_1,\Sbb^T)\nonumber\\
	& \quad  \quad  \quad  +H(Y_{3P}^M|Y_{3D}^M,Y_1^T,Y_2^T,W_1,\Sbb^T)+6T\epsilon \label{eq1}.
}
By using the inequality of Lemma \ref{eq:doubling MI}:
\ea{
	H(Y_{3D}^M|Y_1^T,Y_2^T,W_1,\Sbb^T)& \leq H(Y_1^T|Y_2^T,W_1,\Sbb^T)\nonumber\\
	&  \leq H(Y_1^T|W_1,\Sbb^T)\label{eq:b1},
}
where \eqref{eq:b1} follows due to the fact that conditioning reduces entropy.
Now,  combining applying \eqref{eq:b1} in \eqref{eq1} we conclude that:
\eas{
& T(R_1 + R_2 +R_3 ) \\
& \quad  \quad    \leq H(Y_1^T|\Sbb^T)-H(Y_1^T|W_1,\Sbb^T)\nonumber\\
& \quad  \quad  \quad  +H(Y_2^T|Y_1^T,\Sbb^T)-H(Y_2^T|W_2,Y_1^T,\Sbb^T)\nonumber\\
& \quad  \quad  \quad  +H(Y_{3P}^T|Y_{3D}^T,Y_1^T,Y_2^T,W_1,\Sbb^T)\nonumber\\
& \quad  \quad  \quad   +H(Y_1^T|W_1,\Sbb^T)+6T\epsilon\\
& \quad  \quad   \leq H(Y_1^T|\Sbb^T)+H(Y_2^T|Y_1^T,\Sbb^T)\nonumber\\
& \quad  \quad  \quad  +H(Y_{3P}^M|Y_{3D}^M,Y_1^T,Y_2^T,W_1,\Sbb^T)+6T\epsilon,
}
Finally, using the fact that $H(Y_i^T|\Sbb^T)\leq T \log P$, we have
\ea{
	&	T(R_1 + R_2 +R_3 )\\
	& \quad  \quad \leq H(Y_1^T|\Sbb^T)+H(Y_2^T|\Sbb^T)+H(Y_{3P}^M|\Sbb^T)\\
	& \quad  \quad \leq T\log P+ T\log P+ \f T 2 \log P +6T\epsilon \\
	& \quad  \quad =\f 5 2 T \log P +6T\epsilon,
}
so that, by dividing both sides on $ T\log P $ and letting $P \rightarrow\infty $ and $ T \rightarrow\infty $, we concluded that $SDoF_{\rm sum}\leq \dfrac{5}{2}$.
\subsection{K-User $P^K/DP^{K-1}$  BCCM Channel}
\label{Sec. K-User $D1/P1$  BCCM Channel Case}
As per usual, the proof of Th. \ref{th:main theorem 1} is presented  into two parts: the achievability and the converse.
\subsubsection{Achievability}
In the following, we derive an achievable scheme that attains $SDoF_{\rm sum}= \frac{2K-1}{2} $ by having the TX sends messages $W_1,W_2,\dots,W_K $ to the $K$ RXs while attaining secrecy conditions \eqref {confidentiality}.
%
%
This scheme is described as follows. The TX sends symbol $ a _{1} $  to the first receiver as the message $W _{1}$  in the period of $t_P$, and for any other RX such as $k \in [2:K]$, it sends  the tuple symbols $ \left(a_k,b_k \right) $ as message $W_k$ during the two time slots $t_P$ and $t_D$ .
In the time slot $t_P$ that the channel is in state $ P^K $, symbol $ a _{1} $ is sent to the first receiver, and symbol $ a _{k} $ is sent to every RX $k$ such that $k \in [2:K] $. Since in this time slot perfect CSI of all RXs are known at the TX, it is capable of sending the symbols at any direction knowing that the irrelevant RXs will not conceive anything about the symbol. Beside, the desired RX privately receives each symbol via using the coefficients of channel states and choosing the suitable direction of interference beamforming vectors. We show interference beamforming vectors with $ {\vv}_1(t_P),{\vv}_2(t_P),\dots,{\vv}_K(t_P) $ such that $\forall k \in [K]$, $ {\vv}_k(t_P) $ is a normalized column vector of order $ K $ for $k^{\rm th}$ RX ($ {\vv}_k(t_P) \in \Cbb^{K} $). In the time slot $t_P$, according to \eqref{I/O} the input/output equations are:
\eas{
	X(t_P) &= a_{k}{\vv}_k(t_P)+\sum_{i\in [K]
		,i\neq k}a_{i}{\vv}_i(t_P)\label{3.1}\\
	\forall k \in [K],\,
	Y_k(t_P)& = a_{k}\left\langle {\hv}_k(t_P),{\vv}_k(t_P)\right\rangle  +
	\underbrace{\sum_{i\in [K],i\neq k}a_{i}\left\langle {\hv}_k(t_P),{\vv}_i(t_P) \right\rangle }_\text{interference}\label{3.2}.
}
where the latter term in \eqref{3.2} represents interference at RX $k$ so that these signals are not carrying useful information for this RX.
Note that due to high signal to noise ratio, the additive noise is omitted from the output equations. For every receiver $ k $, interference will be completely removed only if the other RXs cannot access the direction of interference beamforming vector $ {\vv}_k(t_P) $. In other words, the $ \left\langle {\hv}_k(t_P),{\vv}_i (t_P)\right\rangle  $ should be zero in equation \eqref{3.2} for every $ k\neq i $ that $  i\in [K]\ $. If we define $ {{\bf{H}}^{\mathsf{H}}}^{/k}(t) $ as a $ (K-1) \times K $ matrix that has been achieved by removing the  $k^{th}$ row from of the channel state matrix ${\bf{H}}^{\mathsf{H}}(t)$ at time instance t, by writing all of the interference effect omitting equations for all of the $K$ RXs, we can calculate interference beamforming vectors as follows:
\label{3.3}
\begin{equation}
	\begin{cases}
		{{\bf{H}}^{\mathsf{H}}}^{/k}(t_P) {\vv}_k(t_P) =0&  \\
		\| {\vv}_k(t_P) \|^2 =1.&
	\end{cases}
\end{equation}
As a result, each RX by using its channel state vector and interference beamforming vector can decode its desired symbols.\\
In the other time slot $t_D$, we assume the channel is in state $ DP^{K-1} $. Since the TX does not know anything about the CSI of RX $1$, there is no possibility to transmit any symbol for this RX confidentially. Besides, we should find a way that this partial CSIT does not have any effect on the other RXs. To solve this problem, we use the technique of transmitting artificial noise $ u\sim \mathcal{C}\mathcal{N}(0, P)$. If we denote the interference beamforming vectors in the second time slot by $ \vv_1(t_D),\vv_2(t_D),\dots,{\vv}_K(t_D) $, the TX at this stage intends to send symbol $b _{k}$ as the other part of  message $W _{k}$ for every receiver $ k $ that $k \in [2:K] $ and $ k\neq1 $. For the input we can write:
\ea{
	\label{3.4}
	X(t_D)= u \vv_1(t_D) +\sum_{i=2}^{K} b_i {\vv}_i(t_D).
}
For each receiver $  k \in [K]\  , k\neq 1 $, the output is:
\ea{
	\label{3.5}
	Y_{k}(t_D)= b_{k}\left\langle {\hv}_{k}(t_D),{\vv}_k(t_D)\right\rangle  +
	\underbrace{u \left\langle {\hv}_k(t_D),\vv_1(t_D)\right\rangle  +\sum_{i=2,i\neq k}^{K}b_{i}\left\langle {\hv}_{k}(t_D),{\vv}_i(t_D)\right\rangle }_\text{interference}.
}
If we consider $ {{\bf{{H}}}^{\mathsf{H}}}^{/k,1}(t) $ as a $ (K-2) \times K $ matrix that has been achieved by removing the first and $k^{\rm th}$ rows of the channel state matrix at time instance t, $ {{\bf{{H}}}^{\mathsf{H}}}(t) $, similar to the previous time slot, to dispose of interference, and deliver the desired symbols to each receiver except the RX $1$ we will use the following:
\begin{equation}\label{3.6}
	\begin{cases}
		{{\bf{H}}^{\mathsf{H}}}^{/k,1}(t_D){\vv}_k(t_D)=0 & \\
		\| {\vv}_k(t_D)\|^2 =1& 
	\end{cases}
\end{equation}
Since for RX $2$ to $K$, perfect CSIT is available; thus, we choose ${\vv}_1(t_D)$ in a way that it be orthogonal for all these channel coefficient vectors. that means we choose ${\vv}_1(t_D)$ as ${{\bf{H}}^{\mathsf{H}}}^{/1}(t_D){\vv}_1(t_D)=0$ . For the first receiver, the output will be calculated as follows:
\eas{
	\label{3.7}
	Y_{1}(t_D)&=  u\left\langle  {\hv}_1{(t_D)}, \vv_1(t_D)\right\rangle +\sum_{i=2}^{K} b_i\left\langle  {\hv}_1(t_D), {\vv}_i(t_D)\right\rangle \\
	&\overset{\vartriangle}{=}
	L(b_2,\dots,b_K,u).
	\label{eq:te}
}
In equation \eqref{eq:te}, $ L(b_2,\dots,b_K,u) $,  denotes a linear combination of the symbols $ b_2,b_3,\dots,b_K $ , and the artificial noise $ u $. The first RX receives the linear combination with a Gaussian noise  $ u $, and will be unable retrieve irrelevant symbols. In other words, the transmitter sends the high-power Gaussian noise $ u $ in the direction of   ${\vv}_1(t_D)$ so that the unawareness of CSI of this RX would not affect the output of other RXs.\\
In such achievable schema, transmitter sends one symbol to the first RX, and sends two symbols to every other RXs in two time slots that results in the sum SDoF being $ (K-1)+\frac{1}{2} =\frac{2K-1}{2} $.
Now we prove the secrecy constraints for this achievable schema. According to equation \eqref{confidentiality} for each RX $ i $, the secrecy is preserved if the expression $ I(W_i;\lcb Y_{j}^{T} \rcb_{j \in [0:i-1]} |\Scal^T)\leq T\epsilon $ holds. By replacing the outputs of the two time slots, we will have the following for each $ i \in [K]$:
\eas{
	&I(W_i;\lcb Y_{j}^{T} \rcb_{j \in [0:i-1]}|\Scal^T) \nonumber \\
	&= I(a_i,b_i;\lcb a_k\left\langle {\hv}_k(t_P),{\vv}_k(t_P)\right\rangle   \rcb_{k \in [0:i-1]},L(b_2,\dots,b_K,u),\lcb b_r \left\langle {\hv}_r(t_D),{\vv}_r(t_D)\right\rangle  \rcb_{r \in [0:i-1]}|\Scal^T) \nonumber \\
	&= I(a_i,b_i;L(b_2,\dots,b_K,u)|\Scal^T) \nonumber \\
	&{=} I(b_i;L(b_2,\dots,b_K,u)|\Scal^T)+
	I(a_i;L(b_2,\dots,b_K,u)|\Scal^T)\label{ak1}  \\
	&{\leq}\Ocal(\log P),\label{ak2}
}
where \eqref{ak1} results form independence of messages, and \eqref{ak2} results from the fact that due to the high power of the artificial noise, it is impossible to attain the linear combination of desired symbols.
\subsubsection{Converse}
We first introduce a property which will be useful to establish the results in this work and is called statistical equivalence property (SEP) \cite{awan2016sdof} and then by using this lemma we presented the proof of converse part in Appendix \ref{k-converse}.\\
Consider the channel input/output relationship \eqref{I/O} for RX $1$ at the channel state $DP^{K-1} $:
\ea{\label{conv1}
	Y_{1,DP^{K-1}}(t)={\hv}_{1,DP^{K-1}}^{\mathsf{H}}(t)X(t)+N_{1,DP^{K-1}}(t)
}
We want to define a virtual RX or a statistically indistinguishable RX for the first RX in a way that the channel output for this RX is independent of the channel output of the actual RX, and its distribution is the same as the channel output of the actual RX. Therefore, $ {\hv}_{1,DP^{K-1}}^{\mathsf{H}}(t) $ is replaceable with $ \tilde{\hv}_{1,DP^{K-1}}^{\mathsf{H}}(t)$ in the virtual RX in a way that these two vectors are independent, and their distributions are the same. Likewise, we can replace Gaussian noise $ N_{1, DP^{K-1}}(t) $  with $ \tilde{N}_{1, DP^{K-1}}(t) $ such that they have independent identical distribution. Given these, the output of the virtual RX is given by:
\ea{\label{conv2}
	\Yt_{1,DP^{K-1}}(t)=\tilde{\hv}_{1,DP^{K-1}}^{\mathsf{H}}(t)X(t)+\tilde{N}_{1,DP^{K-1}}(t).
}
\begin{lem}{\bf $K$-user SEP.}
	\label{lem_sep1}
	If $\Scal^T=\lcb \Hv(i) \rcb_{i \in [T]},$ $\tilde{\Scal}^T=\lcb \tilde{\Hv}(i) \rcb_{i \in [T]}$, $\Sbb^T=\left\lbrace \Scal^T, \tilde{\Scal}^T\right\rbrace$ and $\Omega $ is an auxiliary random variable, then, the SEP states that:
	\ea{
		H(\tilde{Y}_{1,DP^{K-1}}(t)|
		{Y}_{1,DP^{K-1}}^{t-1},\Omega,\Sbb^T)=
		H({Y}_{1,DP^{K-1}}(t)|
		{Y}_{1,DP^{K-1}}^{t-1},\Omega,\Sbb^T)
	}
\end{lem}
\begin{lem}
	\label{lem_sep2}
	For the channel model with the secrecy conditions \eqref{I/O} with alternating CSIT between the states  $ {P^K} $ and ${DP^{K-1}}$, we have:
	\ea{
		\label{sep2}
		&2H({Y}_{1}^{T}| \lcb Y_{j}^{T} \rcb_{j \in [0:K-1],j\neq 1} , W_{1},\Sbb^T)  \geq
		H({Y}_{1}^{T},{Y}_{K,DP^{K-1}}^{M}|\lcb Y_{j}^{T} \rcb_{j \in [0:K-1],j\neq 1}, W_{1},\Sbb^T).
	}	
\end{lem}
	The proof is given in Appendix \ref{app:lem_sep2}.
\end{IEEEproof}
\section{CONCLUSION}
In this paper, we used synergistic benefits of alternating CSIT to study SDoF of a  $K$-user  Multiple Input/Single Output (MISO)  Broadcast Channel with Confidential Messages (BCCM) and alternating Channel State Information at the Transmitter (CSIT). 
In the MISO BCCM, a transmitter (TX) with $K$ antennas transmit toward $K$ receivers (RXs), in such a way that the message for RX $k$ is kept secret from RX $j$ for all $j<k$.  
The channel between the TX and each RX is a fading channel: the CSI is assumed to be known instantaneously  at the transmitter for the receivers $2$ to $K$.
On the other hand,  the CSI of RX $1$ is known at the transmitter (i) instantaneously for half of the time while (ii) with a unit delay for the remainder of the time.
%
%
%
For this channel model, we calculated the high-SNR characterization of the secure capacity of the sum rate in the form of the Secure Degrees of Freedom (SDoF), as $SDoF=(2K-1)/2$.
%
In achievability proof, we use the benefits of artificial noise transmission to retain confidentiality and exploiting orthogonal space.  For the converse proof, we adopt the so-called statistical equivalence property lemma.
\appendices
\section{Proof of Lemma \ref{lem_sep2}}
\label{app:lem_sep2}
To prove Lemma \eqref{lem_sep2}, We use two different explanation of  $ H({Y}_{1}^{T}|\lcb Y_{j}^{T} \rcb_{j \in [0:K-1], j \neq 1}, W_{1},\Sbb^T) $, once for real RX and other time for virtual RX.\\
The first expansion for the real RX is:
\eas{
	&H({Y}_{1}^{T}|\lcb Y_{j}^{T} \rcb_{j \in [0:K-1], j \neq 1}, W_{1},\Sbb^T)\nonumber\\
	& =H({Y}_{1,DP^{K-1}}^{M},{Y}_{1,P^K}^{M}|\lcb Y_{j}^{T} \rcb_{j \in [0:K-1], j \neq 1}, W_{1},\Sbb^T)\\
	&{=}\sum_{t=1}^{T}H({Y}_{1,DP^{K-1}}(t)
	|Y_{1,DP^{K-1}}^{t-1},\lcb Y_{j}^{T} \rcb_{j \in [0:K-1], j \neq 1}, W_{1},\Sbb^T)\nonumber\\
	& +H({Y}_{1,P^K}^{M}|{Y}_{1,DP^{K-1}}^{M},\lcb Y_{j}^{T} \rcb_{j \in [0:K-1], j \neq 1}, W_{1},\Sbb^T).\label{app1}
}
The second expansion for the virtual RX is:
\eas{
	&H({Y}_{1}^{T}|\lcb Y_{j}^{T} \rcb_{j \in [0:K-1], j \neq 1}, W_{1}, \Sbb^T)\nonumber\\
	& =H({Y}_{1,DP^{K-1}}^{M},{Y}_{1,P^K}^{M}|\lcb Y_{j}^{T} \rcb_{j \in [0:K-1], j \neq 1}, W_{1},\Sbb^T)\\
	&{=}\sum_{t=1}^{T}H(\tilde{Y}_{1,DP^{K-1}}(t)
	|Y_{1,DP^{K-1}}^{t-1},\lcb Y_{j}^{T} \rcb_{j \in [0:K-1], j \neq 1}, W_{1},\Sbb^T)\nonumber\\
	& +H({Y}_{1,P^K}^{M}|{Y}_{1,DP^{K-1}}^{M},\lcb Y_{j}^{T} \rcb_{j \in [0:K-1], j \neq 1}, W_{1},\Sbb^T),\label{app2}
}
where \eqref{app1} and \eqref{app2} follow by Chain rule. Now by using summation of equations \eqref{app1} and \eqref{app2} we can follow:
\eas{
	& 2H({Y}_{1}^{T}|\lcb Y_{j}^{T} \rcb_{j \in [0:K-1], j \neq 1}, W_{1},\Sbb^T)\nonumber\\
	&=\sum_{t=1}^{T}H({Y}_{1,DP^{K-1}}(t)
	|Y_{1,DP^{K-1}}^{t-1},\lcb Y_{j}^{T} \rcb_{j \in [0:K-1], j \neq 1}, W_{1},\Sbb^T)\nonumber\\& +
	\sum_{t=1}^{T}H({\tilde{Y}}_{1,DP^{K-1}}(t)
	|{Y}_{1,DP^{K-1}}^{t-1},\lcb Y_{j}^{T} \rcb_{j \in [0:K-1], j \neq 1}, W_{1},\Sbb^T)\nonumber\\
	& + 2H({Y}_{1,P^K}^{M}|{Y}_{1,DP^{K-1}}^{M},\lcb Y_{j}^{T} \rcb_{j \in[0:K-1], j \neq 1}, W_{1},\Sbb^T)\\
	&{\geq}\sum_{t=1}^{T}H({Y}_{1,DP^{K-1}}(t)
	|{Y}_{1,DP^{K-1}}^{t-1},\lcb Y_{j}^{T} \rcb_{j \in [0:K-1], j \neq 1}, W_{1},\Sbb^T)\nonumber\\& +
	\sum_{t=1}^{T}H({\tilde{Y}}_{1,DP^{K-1}}(t)
	|{Y}_{1,DP^{K-1}}^{t-1},\lcb Y_{j}^{T} \rcb_{j \in [0:K-1], j \neq 1}, W_{1},\Sbb^T)\nonumber\\
	&+H({Y}_{1,P^K}^{M}|{Y}_{1,DP^{K-1}}^{M},\lcb Y_{j}^{T} \rcb_{j \in [0:K-1], j \neq 1}, W_{1},\Sbb^T)\nonumber\\
	&+T\Ocal(\log P)\label{app3}\\
	&\geq \sum_{t=1}^{T}H({{Y}_{1,DP^{K-1}}(t),\tilde{Y}}_{1, DP^{K-1}}(t)|{Y}_{1,DP^{K-1}}^{t-1}, \lcb Y_{j}^{T} \rcb_{j \in [0:K-1], j \neq 1}, W_{1},\Sbb^T)\nonumber\\&+ H({Y}_{1,P^K}^{T}|{Y}_{1,DP^{K-1}}^{T},\lcb Y_{j}^{T} \rcb_{j \in [0:K-1], j \neq 1}, W_{1},\Sbb^T)+T\Ocal(\log P)\\
	& = \sum_{t=1}^{T}H({{Y}_{1,DP^{K-1}}(t),\tilde{Y}}_{1, DP^{K-1}}(t),{Y}_{K,DP^{K-1}}(t)
	|{Y}_{1,DP^{K-1}}^{t-1},\lcb Y_{j}^{T} \rcb_{j \in [0:K-1], j \neq 1}, W_{1},\Sbb^T)
	\nonumber\\&-\sum_{t=1}^{T} H({Y}_{K,DP^{K-1}}(t)|
	{Y}_{1,DP^{K-1}}(t),\tilde{Y}_{1,DP^{K-1}}(t),{Y}_{1,DP^{K-1}}^{t-1},\lcb Y_{j}^{T} \rcb_{j \in [0:K-1], j \neq 1}, W_{1},\Sbb^T) \nonumber\\
	& + H({Y}_{1,P^K}^{M}|{Y}_{1,DP^{K-1}}^{M},\lcb Y_{j}^{T} \rcb_{j \in [0:K-1], j \neq 1}, W_{1},\Sbb^T)+T\Ocal(\log P),\label{step1}
}
where \eqref{app3} follows by:
\begin{align}
	&H({Y}_{1,P^K}^{M}|{Y}_{1,DP^{K-1}}^{M},\lcb Y_{j}^{T} \rcb_{j \in [0:K-1], j \neq 1}, W_{1},\Sbb^T)\nonumber\\&\geq
	H({Y}_{1,P^K}^{M}|{Y}_{1,DP^{K-1}}^{M},\lcb Y_{j}^{T} \rcb_{j \in [0:K-1], j \neq 1}, W_{1},X^T,\Sbb^T)\nonumber\\&=T\Ocal(\log P).
\end{align}
We continue to lower bound \eqref{step1} as the following:
\eas{
	& 2H({Y}_{1}^{T}|\lcb Y_{j}^{T} \rcb_{j \in [0:K-1], j \neq 1}, W_{1},\Sbb^T)\nonumber\\
	& {\geq}
	\sum_{t=1}^{T}H({{Y}_{1,DP^{K-1}}(t),\tilde{Y}}_{1, DP^{K-1}}(t),{Y}_{K,DP^{K-1}}(t)|{Y}_{1,DP^{K-1}}^{t-1},\lcb Y_{j}^{T} \rcb_{j \in[0:K-1], j \neq 1}, W_{1},\Sbb^T)\nonumber\\
	&+ H({Y}_{1,P^K}^{M}|{Y}_{1,DP^{K-1}}^{M},\lcb Y_{j}^{T} \rcb_{j \in [0:K-1], j \neq 1}, W_{1},\Sbb^T)+T\Ocal(\log P)\label{app4}\\
	& {\geq}\sum_{t=1}^{T}H({{Y}_{1,DP^{K-1}}(t),\tilde{Y}}_{1,DP^{K-1}}(t)
	,{Y}_{K,DP^{K-1}}(t)|{Y}_{1,DP^{K-1}}^{t-1},{Y}_{K,DP^{K-1}}^{t-1},\lcb Y_{j}^{T} \rcb_{j \in [0:K-1], j \neq 1}, W_{1},\Sbb^T)\nonumber\\
	&+ H({Y}_{1,P^K}^{M}|{Y}_{1,DP^{K-1}}^{M},\lcb Y_{j}^{T} \rcb_{j \in [0:K-1], j \neq 1}, W_{1},\Sbb^T)+T\Ocal(\log P)\label{app5}\\
	&\geq H({Y}_{1,DP^{K-1}}^{M},{Y}_{K,DP^{K-1}}^{M}|\lcb Y_{j}^{T} \rcb_{j \in [0:K-1], j \neq 1}, W_{1},\Sbb^T)\nonumber\\
	&+ H({Y}_{1,P^K}^{M}|{Y}_{1,DP^{K-1}}^{M},{Y}_{K,DP^{K-1}}^{T},\lcb Y_{j}^{T} \rcb_{j \in [0:K-1], j \neq 1}, W_{1},\Sbb^T)+T\Ocal(\log P)\\
	&=H({Y}_{1,P^K}^{M},{Y}_{1,DP^{K-1}}^{M},{Y}_{K,DP^{K-1} }^{T}|\lcb Y_{j}^{T} \rcb_{j \in [0:K-1], j \neq 1}, W_{1},\Sbb^T)
	+T\Ocal(\log P)\nonumber\\
	&=H({Y}_{K,DP^{K-1}}^{M},{Y}_{1}^{T}|
	\lcb Y_{j}^{T} \rcb_{j \in [0:K-1], j \neq 1}, W_{1},\Sbb^T)+T\Ocal(\log P),
}
where \eqref{app4} follows because given $$ \left({Y}_{1,DP^{K-1}}(t),\tilde{Y}_{1,DP^{K-1}}(t),{Y}_{1,DP^{K-1}}^{t-1},\lcb Y_{j}^{T} \rcb_{j \in [0:K-1], j \neq 1} \right), $$ we can
reconstruct $ {Y}_{K,DP^{K-1}}(t) $ within noise distortion and
\eqref{app5} follow from the fact that conditioning reduces entropy. \\
so we can follow as:
\ea{
	&2H({Y}_{1}^{T}|\lcb Y_{j}^{T} \rcb_{j \in [0:K-1], j \neq 1}, W_{1},\Sbb^T)\nonumber\\
	&  \geq H({Y}_{K,DP^{K-1}}^{M},{Y}_{1}^{T}|
	\lcb Y_{j}^{T} \rcb_{j \in [0:K-1], j \neq 1}, W_{1},\Sbb^T)\\
	&= H({Y}_{1}^{T}|\lcb Y_{j}^{T} \rcb_{j \in [0:K-1], j \neq 1}, W_{1},\Sbb^T)\nonumber\\&+H({Y}_{K,DP^{K-1}}^{M}|\lcb Y_{j}^{T} \rcb_{j \in [0:K-1], j \neq 1},{Y}_{1}^{T}, W_{1},\Sbb^T).
}
This conlcudes the proof of lemma.
\section{Proof of Converse for $K$-user}
\label{k-converse}
We begin the converse proof as follows:
\eas{
	\sum_{i=1}^{K}T R_i & {=}\sum_{i=1}^{K-1}H(W_i|\Sbb^T)+H(W_K|W_{1},\Sbb^T)\label{cvp1}\\
	&=\sum_{i=1}^{K-1}I(W_i;Y_{i}^{T}|\Sbb^T)+H(W_i|Y_{i}^{T},\Sbb^T)
	+ I(W_K;Y_{K}^{T}|W_{1},\Sbb^T)+H(W_K|Y_{K}^{T},W_{1},\Sbb^T)\\
	&\leq\sum_{i=1}^{K-1}I(W_i;Y_{i}^{T}|\Sbb^T)+I(W_K;Y_{K}^{T}|
	W_{1},\Sbb^T)+ KT\epsilon\label{cvp2}\\
	&{\leq} I(W_1;Y_1^T|\Sbb^T)+\sum_{i=2}^{K-1}I(W_i;Y_i^T|\lcb Y_{j}^{T} \rcb_{j \in [0:i-1]},\Sbb^T)
	+ I(W_{K};Y_{K}^T|W_{1},\lcb Y_{j}^{T} \rcb_{j \in [0:K-1]},\Sbb^T)+ 2KT\epsilon\label{cvp3} \\
	&=\sum_{i=1}^{K-1} H(Y_i^T|\lcb Y_{j}^{T} \rcb_{j \in [0:i-1]},\Sbb^T) 
	-H(Y_i^T|\lcb Y_{j}^{T} \rcb_{j \in [0:i-1]},W_i,\Sbb^T) \nonumber \\
	& \quad +H(Y_K^T|\lcb Y_{j}^{T} \rcb_{j \in [0:K-1]},W_{1},\Sbb^T)-
	H(Y_K^T|\lcb Y_{j}^{T} \rcb_{j \in [0:K-1]},W_{1},W_K,\Sbb^T)+2TK\epsilon,\label{step2}
}
where \eqref{cvp1} follows by using the independency of messages from each other and channel states, \eqref{cvp2} follows by Fano’s inequality, \eqref{cvp3} follows from \eqref{cvp4} and \eqref{cvp5} in the following. For RXs one to $K-1$ we can obtain:
\eas{
	I(W_i;Y_{i}^{T}|\Sbb^T) & \leq I(W_i;\lcb Y_{j}^{T} \rcb_{j \in [0:i]}|\Sbb^T)\\
	&=I(W_i;\lcb Y_{j}^{T} \rcb_{j \in [0:i-1]}|\Sbb^T)+I(W_i;Y_i^T|\lcb Y_{j}^{T} \rcb_{j \in [0:i-1]},\Sbb^T)\\
	&\leq
	I(W_i;Y_i^T|\lcb Y_{j}^{T} \rcb_{j \in [0:i-1]},\Sbb^T) +T\epsilon. \label{cvp4}
}
For RX $K$, we obtained a little bit different inequality which follows:
\eas{
	&I(W_K;Y_{K}^{T}|W_{1},\Sbb^T)\nonumber\\&\leq I(W_K;\lcb Y_{j}^{T} \rcb_{j \in [0:K]}|W_{1},\Sbb^T)\\
	& =I(W_K;W_{1},Y_K^T,\lcb Y_{j}^{T} \rcb_{j \in [0:K-1]}|\Sbb^T)\label{cvp4_1}\\
	&= I(W_K;\lcb Y_{j}^{T} \rcb_{j \in [0:K-1]}|\Sbb^T)+
	I(W_{K};W_{1}|\lcb Y_{j}^{T} \rcb_{j \in [0:K-1]},\Sbb^T)\nonumber\\
	&\quad+I(W_{K};Y_{K}^T|W_{1},\lcb Y_{j}^{T} \rcb_{j \in [0:K-1]},\Sbb^T)\\
	& \leq I(W_{K};Y_{K}^T|W_{1},\lcb Y_{j}^{T} \rcb_{j \in [0:K-1]},\Sbb^T)
	+H(W_{1}|\lcb Y_{j}^{T} \rcb_{j \in [0:K-1]},\Sbb^T)\nonumber\\
	& \quad -H(W_{1}|W_{K},\lcb Y_{j}^{T} \rcb_{j \in [0:K-1]},\Sbb^T)+
	T\epsilon \label{cvp4_2} \\
	&\leq I(W_{K};Y_{K}^T|W_{1},\lcb Y_{j}^{T} \rcb_{j \in [0:K-1]},\Sbb^T)+ H(W_{1}|\lcb Y_{j}^{T} \rcb_{j \in [0:K-1]},\Sbb^T)+T\epsilon\\
	& \leq I(W_{K};Y_{K}^T|W_{1},\lcb Y_{j}^{T} \rcb_{j \in [0:K-1]},\Sbb^T)+H(W_{1}|Y_{1}^T,\Sbb^T)+T\epsilon\\
	& \leq I(W_{K};Y_{K}^T|W_{1},\lcb Y_{j}^{T} \rcb_{j \in [0:K-1]},\Sbb^T) +2T\epsilon\label{cvp5},
}
where \eqref{cvp4} and \eqref{cvp4_2} follows from the secrecy conditions, \eqref{cvp4_1} follows with independence of messages and, \eqref{cvp5} follows by Fano’s inequality. We continue to upper bound \eqref{step2} in the following:
\eas{
	\sum_{i=1}^{K}TR_i & \leq\sum_{i=1}^{K-1}  H(Y_i^T|\lcb Y_{j}^{T} \rcb_{j \in [0:i-1]},\Sbb^T)
	-H(Y_i^T|\lcb Y_{j}^{T} \rcb_{j \in [0:i-1]},W_i,\Sbb^T) \nonumber \\
	& +H(Y_K^T|\lcb Y_{j}^{T} \rcb_{j \in [0:K-1]},W_{1},\Sbb^T) 
	-H(Y_K^T|\lcb Y_{j}^{T} \rcb_{j \in [0:K-1]},W_{1},W_K,\Sbb^T)+2TK\epsilon\\
	&= \sum_{i=1}^{K-1}  H(Y_i^T|\lcb Y_{j}^{T} \rcb_{j \in [0:i-1]},\Sbb^T) -
	H(Y_i^T|\lcb Y_{j}^{T} \rcb_{j \in [0:i-1]},W_i,\Sbb^T) \nonumber\\
	&+H({Y}_{K,DP^{K-1}}^{T},{Y}_{K,P^K}^{T}|\lcb Y_{j}^{T} \rcb_{j \in [0:K-1]},W_{1},\Sbb^T)-
	H({Y}_{K}^T|\lcb Y_{j}^{T} \rcb_{j \in [0:K-1]},W_{K},W_{1},\Sbb^T)+2TK\epsilon\\
	& \leq \sum_{i=2}^{K-1}H(Y_i^T|\lcb Y_{j}^{T} \rcb_{j \in [0:i-1]},\Sbb^T)
	-H(Y_i^T|\lcb Y_{j}^{T} \rcb_{j \in [0:i-1]},W_i,\Sbb^T)+H(Y_{1}^T|\Sbb^T) \nonumber\\
	&\quad -H(Y_{1}^T|W_{1},\Sbb^T)
	+ H({Y}_{K, DP^{K-1}}^{M}|\lcb Y_{j}^{T} \rcb_{j \in [0:K-1]},W_{1},\Sbb^T)+
	H({Y}_{K,P^K}^{M}|{Y}_{K,DP^{K-1}}^{M},\lcb Y_{j}^{T} \rcb_{j \in [0:K-1]},W_{1},\Sbb^T)\nonumber\\
	&\quad\quad- H({Y}_{K}^{T}|\lcb Y_{j}^{T} \rcb_{j \in [0:K-1]},W_{K},W_{1},\Sbb^T)
	+2TK\epsilon\label{cvp6}
}

by using Lemma \ref{lem_sep2} we can write:
\eas{
	H({Y}_{K,DP^{K-1}}^{T}|\lcb Y_{j}^{T} \rcb_{j \in [0:K-1],j\neq 1},Y_{1}^{T}, W_{1},\Sbb^T)& \leq
	H({Y}_{1}^{T}| \lcb Y_{j}^{T} \rcb_{j \in [0:K-1],j\neq 1} , W_{1},\Sbb^T)\\
	&\overset{(a)}{\leq} H({Y}_{1}^{T}| W_{1},\Sbb^T), \label{cvp7}
}
where \eqref{cvp7} follow from the fact that conditioning reduces entropy.	Now by applying \eqref{cvp7} in \eqref{cvp6}, we get:
\ea{
	\sum_{i=1}^{K}TR_i& \leq \sum_{i=2}^{K-1}  H(Y_i^T|\lcb Y_{j}^{T} \rcb_{j \in [0:i-1]},\Sbb^T) -
	\sum_{i=2}^{K-1}H(Y_i^T|\lcb Y_{j}^{T} \rcb_{j \in [0:i-1]},W_i,\Sbb^T)\nonumber\\
	& \quad + H(Y_{1}^T|\Sbb^T)- H(Y_{1}^T|W_{1},\Sbb^T)
	+H(Y_{1}^T|W_{1},\Sbb^T)
	+H({Y}_{K,P^K}^{M}|{Y}_{K,DP^{K-1}}^{M},\lcb Y_{j}^{T} \rcb_{j \in [0:K-1]},W_{1},\Sbb^T)\nonumber\\
	& \quad \quad - H({Y}_{K}^{T}|\lcb Y_{j}^{T} \rcb_{j \in [0:K-1]},W_{K},W_{1},\Sbb^T) +2TK\epsilon\\
	&=\sum_{i=1}^{K-1}  H(Y_i^T|\lcb Y_{j}^{T} \rcb_{j \in [0:i-1]},\Sbb^T)+
	H({Y}_{K,P^K}^{M}|{Y}_{K,DP^{K-1}}^{M},\lcb Y_{j}^{T} \rcb_{j \in [0:K-1]},W_{1},\Sbb^T)\nonumber\\
	& \quad -\sum_{i=2}^{K-1}H(Y_i^T|\lcb Y_{j}^{T} \rcb_{j \in [0:i-1]},\Sbb^T)-H({Y}_{K}^{T}|\lcb Y_{j}^{T} \rcb_{j \in [0:K-1]},W_{K},W_{1},\Sbb^T) +2TK\epsilon\\
	& \leq \sum_{i=1}^{K-1}  H(Y_i^T|\lcb Y_{j}^{T} \rcb_{j \in [0:i-1]},\Sbb^T)+
	H({Y}_{K,P^K}^{T}|{Y}_{K,DP^{K-1}}^{T},\lcb Y_{j}^{T} \rcb_{j \in [0:K-1]},W_{1},\Sbb^T)
	+2TK\epsilon\\
	&\leq(K-1)T\log P + \f T 2 \log P+2TK\epsilon\\
	& =T\log P ((K-1)+\frac{1}{2})+2TK\epsilon. \label{cvp8}\\
}
By dividing both sides on $ T\log P $, letting $P \rightarrow\infty $ we concluded that:
\begin{equation}
	SDoF_{\rm sum}\leq (K-1)+\frac{1}{2}= \dfrac{2K-1}{2}
\end{equation}
\bibliographystyle{IEEEtran} 
\bibliography{main_ref}
\end{document}